\documentclass[journal=nalefd,manuscript=letter,layout=twocolumn]{achemso}
\usepackage{chemformula} 
\usepackage[T1]{fontenc} 
\usepackage{graphicx}    

\usepackage{siunitx}
\usepackage[utf8]{inputenc}
\usepackage{amssymb}

\usepackage{xcolor} 
\usepackage{soul}  
\author{J.~Fekete}
\email{J.Fekete@sussex.ac.uk}
\author{P.~Joshi}
\author{T.~J.~Barrett}
\author{T.~M.~James}
\author{R.~Shah}
\author{A.~Gadge}
\author{S.~Bhumbra}
\author{W.~Evans}
\altaffiliation{Physikalisch-Technische Bundesanstalt, 10587 Berlin, Germany}
\author{F.~Oru{\v c}evi{\' c}}
\author{P.~Kr\"uger}
\affiliation[Sussex]
{Department of Physics and Astronomy, School of Mathematical and Physical Sciences, University of Sussex, Brighton, BN1 9QH, United Kingdom}
\altaffiliation{Physikalisch-Technische Bundesanstalt, 10587 Berlin, Germany}

\title{Quantum gas-enabled direct mapping of active current density in percolating networks of nanowires}

\abbreviations{AFM, C-AFM, BEC, BEC-M, PCB, 1D, 2D}


\begin{document}


\begin{abstract}
    Electrically percolating nanowire networks are amongst the most promising candidates for next-generation transparent electrodes. Scientific interest in these materials stems from their intrinsic current distribution heterogeneity, leading to phenomena like percolating pathway re-routing and localized self-heating, which can cause irreversible damage. Without an experimental technique to resolve the current distribution, and an underpinning nonlinear percolation model, one relies on empirical rules and safety factors to engineer these materials. 
    
   We introduce Bose-Einstein microscopy to address the long-standing problem of imaging active current flow in 2D materials. We report on improvement of the performance of this technique whereby observation of dynamic re-distribution of current pathways becomes feasible. We show how this, combined with existing thermal imaging methods,   eliminates the need for assumptions   between  electrical and thermal properties. This will enable testing and modelling individual junction behaviour and hot-spot formation. Investigating both reversible and irreversible mechanisms will contribute to the advancement of devices with improved  performance and  reliability.
    
\end{abstract}

\paragraph{Introduction}
Over the last few decades percolating networks have attracted much interest, due largely to their increasingly important role in the development of transparent conductors, \cite{Chen2013} flexible thin-film transistors, \cite{Kumar2005} as well as chemical \cite{Sysoev2007} and nanobio \cite{Nair2007} sensors. To characterize the relationship between the microscopic structure and the macroscopic physical properties of these networks, percolation theory is commonly employed. Through a statistical approach, this theory has been successful in describing observed scaling laws, and counterintuitive responses for network properties such as electrical conductance and maximum voltage drop \cite{Pimparkar2007, Zezelj2012, Das2016}. These models play a crucial role in optimising the network's performance for various applications. The tunability of key experimental network parameters can in turn be used to test the theoretical models.

In the study of electrical percolating networks, arguably the biggest remaining challenge is the absence of a non-invasive experimental technique capable of directly unveiling a spatially-resolved current density pattern and its often nonlinear, dynamic changes when an electrical voltage is applied across the network. Such a tool would be advantageous where usual statistical models are inadequate, such as in identifying clusters leading to device failure. When used in combination with existing techniques, it would enable in-depth studies of the interplay between electrical, thermal and surface properties, thereby bridging the gap between macroscopic network properties and microscopic observations.

The absence of a viable direct microscopic current density imaging technique has stimulated alternative, indirect measurements of the current flow. Instruments such as scanning electron microscopes, and atomic force microscopes (AFM) equipped with conductive, Kelvin-probe, or magnetic force modules, provide detailed information on the topography, conductivity and important surface properties \cite{SPM2007}. Recently, thermal imaging has been demonstrated as an efficient tool to record thermal maps during active current flow in a network. \cite{Estrada2011,Maize2015,Sannicolo2016} This method revealed interesting mechanisms in hot spot formation and clustering. For the interpretation of the measurements, the relationship between local current density and temperature relies on assumptions which only hold in the low current density limit. However, the thermal changes are only detectable when current densities are high, primarily due to the limited sensitivity of existing methods. In such high-current density regimes, nonlinear thermo-electric interactions are significant, necessitating the use of more complex models.

Here we introduce a solution to the long-standing problem of current density mapping of random two-dimensional (2D) networks. Our approach exploits the capability of quantum gases, such as Bose-Einstein condensates (BEC), to detect ultra-low magnetic fields. BEC microscopy (BEC-M) \cite{Wildermuth2005a,Yang2017a,Taylor2021} offers a unique combination of microscopic resolution, sub-nanoampere sensitivity, and tunable dynamic range. In addition, it is possible to map active current distributions in a single imaging shot rather than by time-consuming scans. Although these properties make this method an outstanding candidate for exploring the nonlinear phenomena in percolating networks, the required performance levels have not been reached so far.

In this Letter, we analyze the feasibility of BEC-M for the study of electrical percolating networks.
We report on benchmark experiments conducted on a microfabricated planar reference structure, demonstrating a substantial improvement over the previously reported performance. Then, at this performance level we demonstrate the feasibility of the method for current density mapping in percolating networks. This is accomplished by simulating BEC-M data obtained from random nanowire networks with varying wire densities. We also discuss the key advantages over AFM. Furthermore, using a single junction model, we show how BEC-M has the potential to deepen our understanding of the thermo-electric interplay in these networks. This interplay underpins critical phenomena such as self-heating and the reconfiguration of current pathways.\cite{Maize2015,Sannicolo2016}

In the context of the challenge of investigating nonlinear percolating networks \cite{Maize2015,Das2016}, we outline how the technique can be used to study nonlinear phenomena and observe dynamics including reversible as well as irreversible mechanisms (see example shown in Figure \ref{fig:Fig4}a) \cite{Garnett2012,Sannicolo2016,Resende2022}.
Finally, we experimentally demonstrate the suitability of BEC microscopy for nanostructured samples, using a network of carbon nanotubes as an example.

\paragraph{BEC microscopy}
\begin{figure*}[!ht]
    \centering
    \includegraphics[width=0.95\textwidth]{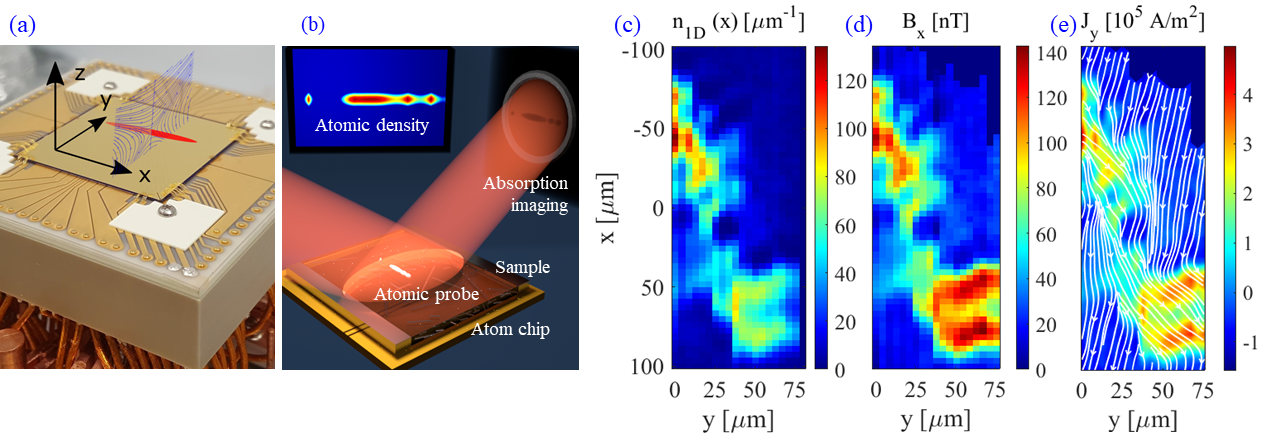}
    \caption{{(a) Trapping structure (PCB and atom chip) and illustration of magnetic field lines in the $y$-$z$ plane of a magnetic quadrupole forming a trap used for the radial confinement of the atomic probe. (b) BEC-M operation scheme. (c) A raster scan of atomic line densities $n_{1\mathrm{D}}(x)$ measured over a planar micro-fabricated conductor. (d) Magnetic field distribution and (e) current density calculated from (d). The streamlines represent current flow, with their density being proportional to the current density $J_x$. To aid the visualisation of the inhomogeneous flow, the current density $J_y$ has been multiplied by a factor of $1000$.}}
    \label{fig:BEC-M}
\end{figure*}
Ever since their theoretical prediction in 1924 \cite{Bose1924} and their first experimental observations in 1995 \cite{Cornell1995}, BECs have garnered significant interest in quantum sensing applications, primarily because they outperform their classical counterparts in numerous aspects \cite{Dickerson2013,Canuel2006,Pandey2019, Wildermuth2005a}.
BEC-M has been introduced \cite{Kruger2005,Wildermuth2005a,Schumm2005,Wildermuth2006,Kruger2007} as a technique to measure the local variation in the magnetic field. The method relies on atom chip technology \cite{Folman2002}, which enables the precise control and manipulation of BECs near surfaces. Small changes in the local magnetic field cause density variations of the atomic gas, which can be directly related back to variations in the current flow. This effect has been measured in lithographically fabricated (evaporated) \cite{Kruger2007} and electroplated wires \cite{Fortagh2002} as well as in microstructured patterns \cite{Yang2017a}. It has also been used to observe the nematic transition in iron pnictide, a high-temperature superconductor \cite{Yang2020}.

The working principle of BEC-M involves neutral atoms in an ultrahigh vacuum (UHV) environment. The atoms are initially laser cooled, then magnetically trapped and further cooled by forced evaporation below the critical temperature for Bose-Einstein condensation (typically hundreds of nanokelvins)\cite{Cornell1995}.
In our experiment, the magnetic trapping relies on precisely controlled currents in various components like atom chips, printed circuit boards (PCB), millimeter-thick wires, and coils to generate a magnetic field with a desired spatial distribution. This flexibility  enables versatile adjustment of the shape and position of the atomic probe. We typically prepare a highly elongated ultracold atomic cloud on an atom chip for microscopy measurements, as depicted schematically in Figure~\ref{fig:BEC-M}a.

For microscopy, a sample of interest is placed between the trapped atoms and the trapping structure. Due to the intrinsically small energy scales in BECs, even tiny magnetic fields emerging from the sample can alter the magnetic trapping potential sufficiently to impact the confinement of the atoms. In response, the atomic wavefunction is distorted and the modified atomic density distribution can be captured using conventional imaging techniques. 

Figure \ref{fig:BEC-M}b shows a schematic illustration of the BEC-M operation. The field distribution emanating from the sample can be reconstructed along the quasi-one dimensional (1D) probe ($x$ direction) from a single shot image taken in a fraction of a millisecond. A 2D-field map can be created by raster scanning the probe position along the orthogonal direction ($y$) across the sample surface.
In our demonstration implementation, each field recording involves the re-creation of the atomic probe, due to the destructive nature of the imaging process, with a duty cycle on the order of seconds.

\paragraph{Current density mapping}
In this section, we demonstrate experimentally an improvement in the state-of-the-art sensitivity of BEC-M, to the point that the method becomes feasible for the application of current density mapping in nanowire networks. For this demonstration, our BECs were scanned over the same microfabricated planar conductor that served as a trapping wire for the atoms, and we show how a current density map is inferred from measured atomic densities (see Supplemental information for further details). The current inhomogeneity observed in this conductor arises from  imperfections in the fabrication process. This measurement provides a benchmark for future measurements on other samples, including nanowire networks.

The atomic densities measured at each position were converted into line densities $n_{1\mathrm{D}}(x)$ resulting in the 2D map shown in Figure 1c. From the line densities, the change in magnetic field component $B_\textrm{x}$, can be approximated\cite{Gerbier2004} by $ 2h \nu_{\perp}a_{sc}/\mu_{\mathrm{B}}\cdot\delta n_{1\mathrm{D}}(x)$, as shown in Figure \ref{fig:BEC-M}d. Here $h$ is Planck's constant, $\nu_{\perp}$ is the radial trapping frequency, $a_{sc}$ is the s-wave scattering length, and $\mu_{\mathrm{B}}$ is the Bohr magneton. The current density $J_y$ (giving rise to the $B_x$ component of the magnetic field) is then calculated from the measured magnetic field distribution using an inverse method,\cite{Roth1989} and is shown in Figure \ref{fig:BEC-M}e.

The single shot noise floor for these measurements was $\qty{0.18}{atoms \per \micro\metre}$, which converts to a minimum detectable field of $\delta B_x = \qty{180}{\pico\tesla}$. The maximum field value of $\qty{140}{\nano\tesla}$ indicates a dynamic range factor of 770. The sensitivity to current is $\qty{1.3}{\nano\ampere}$, assuming an infinitely thin wire as the source of the field. This is a valid approximation in the case of a random nanowire network, as the distance between atoms and the wires is significantly larger than the wire thickness.

Interestingly, we have observed field gradients as large as $\qty{8}{\nano\tesla \per \micro\metre}$ (when averaged over 3 repeats). Such a gradient would allow for probing steps below $\qty{50}{\nano\metre}$ while detecting finite changes in the magnetic field (considering only the field detection limit). However, currently available power supplies used for trapping exceed the noise level that would be required to implement this.

The cloud size (radii of $\qty{290}{\nano\metre}$ radially and $\qty{34}{\micro\metre}$ longitudinally) determines the ultimate spatial resolution and the field-of-view. The resolution of the BEC-M is limited by two parameters. Along $x$, the resolution is limited by optical diffraction, typically a few micrometres. Along $y$, the precise control of the atomic probe position and the radial confinement determine the effective spatial resolution. The orientation of the atomic probe and the scanning direction can be rotated by 90$^\circ$ if high resolution is required in both directions. By tightening the radial confinement (proportional to $1/\sqrt{\nu_{\perp}}$), one can further improve the resolution, however, at the cost of reduced field sensitivity (proportional to $\nu_{\perp}$). 

To put these performance values into context, it is  worth comparing sensitivities of all existing magnetometers. State-of-the-art magnetic field sensors are dominated already by quantum sensors, but there is a trade-off between sensitivity and spatial resolution \cite{Mitchell2020,Yang2017a}. The sensors with highest sensitivity operate in the millimetre to centimetre range, and sensors that can resolve field variation at nanoscale have orders of magnitude lower sensitivity. 
For measuring DC magnetic fields emanating from direct current sources, our improved BEC-M offers the best performance in terms of sensitivity at the microscopic scale.

\begin{figure}
		
		\includegraphics[width=\columnwidth]{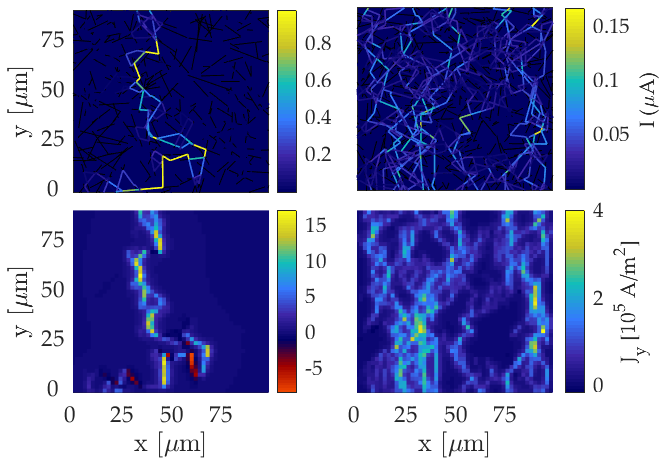}
		\caption{Randomly generated networks of silver nanowires, with 120\,nm wire diameters, subject to $1\, \mu$A input network current. The optical transmittances are 90\% (left) and 80\% (right). Colours indicate the current in each wire (no current in the black wires). Lower panels show the reconstructed current density distributions from the respective networks using BEC-M, for a radial trapping frequency of $\nu_{\perp}=1$\,kHz at an atom-surface distance of $1\,\mu$m. The upper and lower edges of the $\qty{100}{\micro\metre}\times \qty{100}{\micro\metre}$ area represent the two electrodes.}
		\label{fig:Fig2}
\end{figure}

\paragraph{Percolating networks}
We demonstrate the feasibility of the improved BEC-M through simulations on percolating networks of different wire densities, at operating conditions similar to the measurements described in the previous section. We have chosen the parameters from silver nanowires, as they are a key contender for practical applications of the networks. Figure \ref{fig:Fig2} shows randomly generated wire networks (upper panels) and the corresponding reconstructed current density distributions (lower panels) for both a network near the percolation threshold and a high-density network. The network density is related to the transparency of the sample, which is a key quantity in their application for transparent electrodes and touch screens.\cite{Large2016} In the two cases presented here, the optical transmittance values are 90\% and 80\%, respectively.

These examples show that it is possible to distinguish individual current paths and junctions even well above the percolation threshold for standard operating conditions of the BEC-M. One can also observe sections with reverse direction of the current flow, information that other methods are not able to reveal. In the 80\% optical transmittance case hot spots are visible and the current paths can still be resolved.

\paragraph{Method comparison}
\begin{figure*}[!h]
		\includegraphics[width=0.7\textwidth]{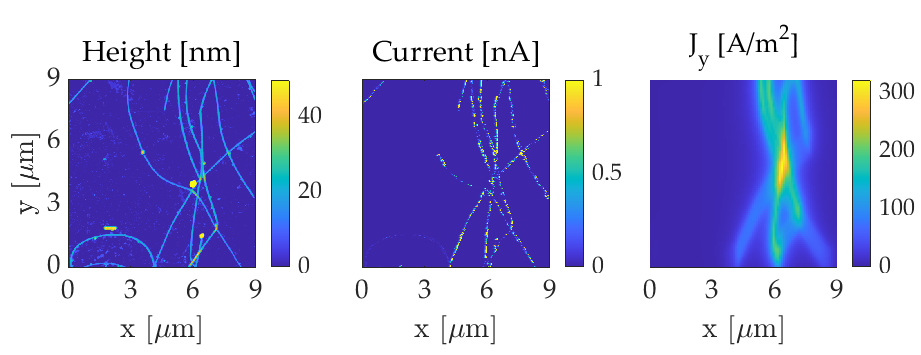}
		\caption{(a) Height map of a  silver nanowire network section measured by AFM and (b) current map recorded by C-AFM. (c) Reconstructed active current density map from the simulation of BECs $\qty{100}{\nano\metre}$ above the network for the same topography for $\qty{100}{\nano\ampere}$ input current.}
		\label{fig:Fig3}
\end{figure*}
To highlight the advantage of BEC-M (i.e. measuring active current flow as opposed to conductivity), we now compare a simulated BEC-M current density map to experimental data measured on a small section of a real nanowire network by atomic force microscopy. The topography scan (Figure~\ref{fig:Fig3}a) and  corresponding conductive (C-AFM) measurement (Figure~\ref{fig:Fig3}b) indicate that these nanowires have approximately $\qty{30}{\nano\metre}$ diameter and typically exceed a length of $\qty{10}{\micro\metre}$. Importantly, one can identify dead-ends that do not carry any current in a live network. These dead-ends appear as conductors, not distinct from the active current-carrying structures, in the C-AFM due to the circuit being closed by the path from the bias electrode to the AFM tip. It can be seen that the measured current value is very sensitive to the tip position relative to the nanowire axes, resulting in a discontinuous map. Therefore, the measured current cannot be expected to be proportional to the current in a live network. Besides this issue, the sensitivity and dynamic range are not appropriate to inform of the presence of hot spots in a network, especially of those forming at low bias voltages.

In contrast to the C-AFM measurement, the current density image obtained from the simulated BEC-M data for the same topography reveals the active current paths only, and shows no false connection at dead-ends. Absolute measurements of the current density can be evaluated if the topography is known, and has a dynamic range at least two orders of magnitude above the minimum detection limit. Figure \ref{fig:Fig3}c was simulated by using the network topography in Figure \ref{fig:Fig3}a with bulk silver resistivity, assuming a typical junction resistance value of $\qty{100}{\ohm}$ and $\qty{100}{\nano\ampere}$ input current applied to the network section (via two electrodes at the top and bottom edges). This represents a realistic example of a section where a hot spot can form within a larger network. The presence of multiple junctions with resistances higher than a nanowire segment of equivalent area, gives rise to the formation of hot spots, as is evident in the simulated current map.

The current is assigned to the network following Kirchoff's law, and produces a magnetic field distribution at the location of the probe, which affects the atomic distribution. The current density is then recovered from the field by an inverse method.\cite{Roth1989} 


\paragraph{Nonlinearity in percolating networks}

We now turn to studying percolating networks in terms of the thermo-electric interplay within and between their components. The microscopic observations of single nanowire and single junction resistances \cite{Bellew2015} as well as those obtained by thermal imaging\cite{Maize2015} still fail to explain the non-Ohmic---also termed as super-Joule---behaviour of junctions, which forms the basis for models aiming to reproduce the dynamical current path redistribution over the network. 
Furthermore, the rich dynamics at microscopic scales are not necessarily reflected in macroscopic observations, as in examples where  substantial reconfiguration of the current does not lead to appreciable change in the total resistance of the network. 

In this section we introduce a model to show how such reconfiguration of the networks can be understood via Joule heating of the wires and a nonlinear junction response to the temperature. The model consists of two wires in contact at a junction (Figure 4b). We assume the junction to have a temperature-dependent resistance, as shown in Figure 4c (inset), exhibiting an abrupt drop in resistance $R_{\mathrm{J}}(T_\mathrm{J})$ at a certain temperature, consistent with the measurements of Bellew et al.\cite{Bellew2015}   
\begin{figure*}[!ht]
		\centering
		\includegraphics[width=1\textwidth]{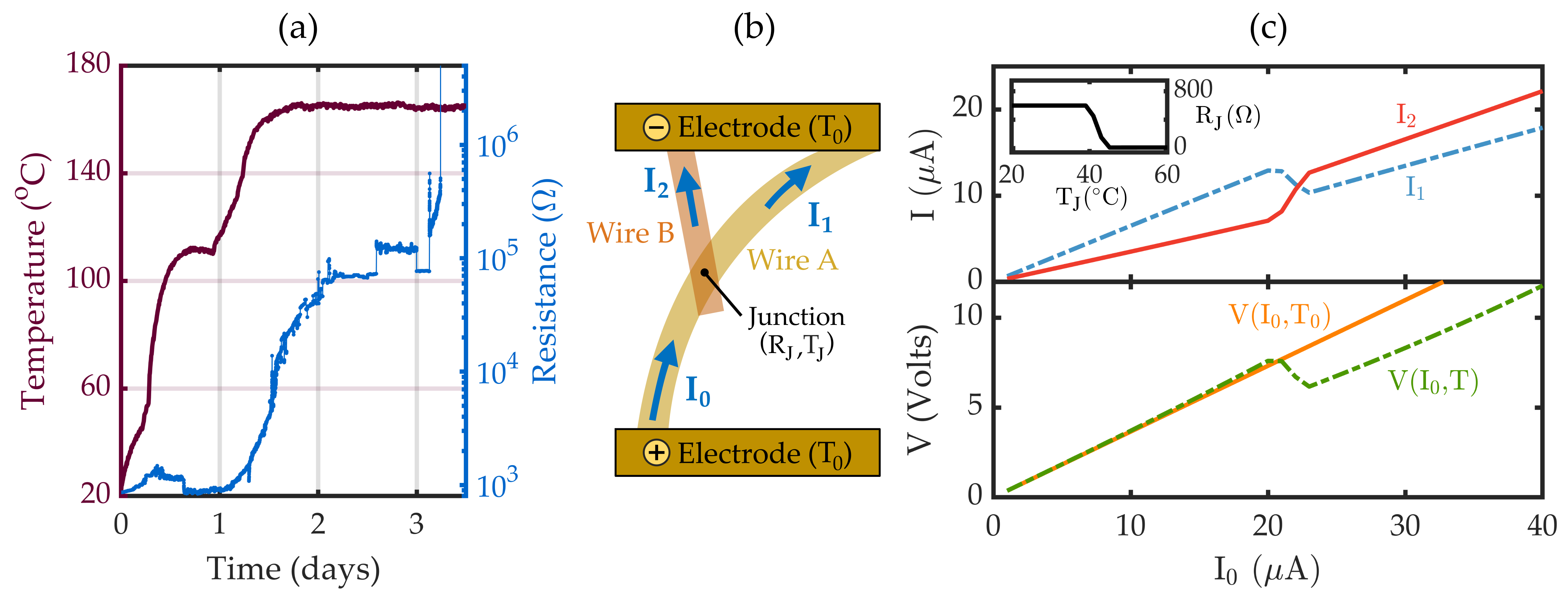}
		\caption{(a) Measurement of a nanowire network resistance while the sample was heated, exhibiting nonlinear response. The abrupt increase in resistance in the final stage is irreversible and corresponds to network failure. (b) Schematic of the model setup, showing two wires crossing at a junction between two electrodes. The junction has resistance $R_\mathrm{J}$ and local temperature $T_\mathrm{J}$. (c) Currents I$_{1}$ and I$_{2}$ (upper plot) in the two upper wire sections of the circuit. Inset depicts the temperature-dependent junction resistance prescribed in the model. Voltage across the circuit (lower plot), showing the I-V characteristic of the circuit when taking into account the temperature dependence in the model $V(I_0,T)$, and when ignoring it $V(I_0,T_0)$. The abrupt change in voltage corresponds to a current path reconfiguration.}
		\label{fig:Fig4}
\end{figure*}

 For a fixed current $I_0$ between the electrodes we calculate the temperature and resistance along the sections of wire and the resulting currents using the 1D heat diffusion equation (see  Supplemental Information for details). For certain parameters we observe current path reconfiguration [Figure~\ref{fig:Fig4}c (upper)] due to the nonlinear junction response, which brings the system into the regime of super-Joule heating. 

The voltage across the circuit [$V(I_0,T)$ in Figure \ref{fig:Fig4}c (lower)] shows a similar behaviour to that observed in Bellew et al. \cite{Bellew2015}. The linear $I$-$V$ regime at low currents and a slight deviation from it at medium current levels is followed by a significant drop, after which linearity is regained at larger currents. Such behaviour is considerably different from the case where temperature dependence is not taken into account [$V(I_0,T_0)$ in Figure \ref{fig:Fig4}c (lower)].

In contrast to thermal maps, which are proportional to the Joule heat, active current maps obtained by BEC-M at various current levels (as low as nanoamperes) enable quantitative analysis of the reconfiguration. The complete information to model percolating networks and work out individual junction behaviour will be available from the combination of the two methods.

\paragraph{BEC-M implementation}

\begin{figure*}[!hbt]
		\centering
		\includegraphics[width=0.85\textwidth]{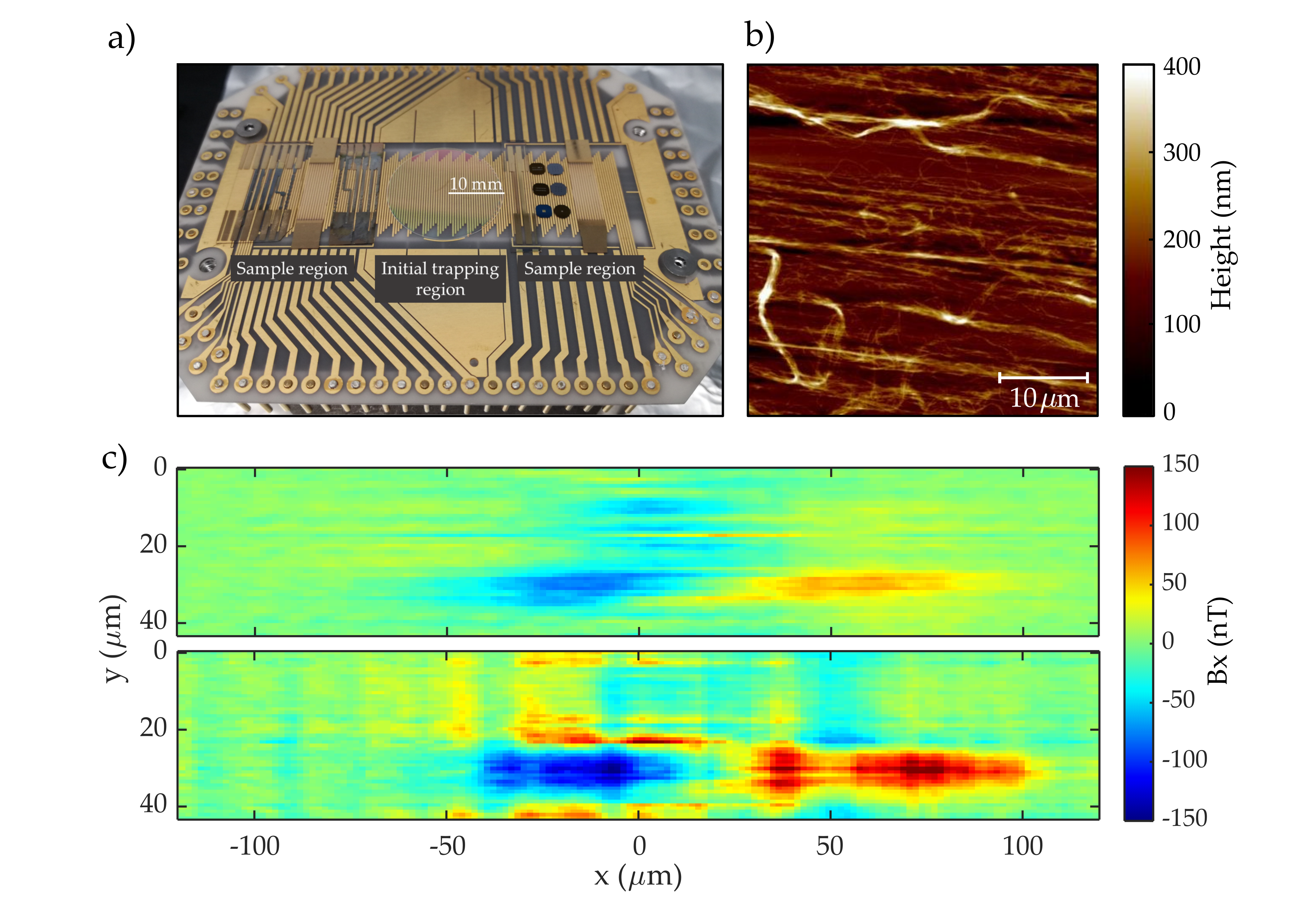}
		\caption{(a) Photograph of a PCB capable of magnetically transporting atoms from the initial trapping and cooling region to the samples, where BEC-M scanning takes place. (b) AFM height map of a representative [not the same as scanned in c)] area of a carbon nanotube sample. The typical features that explain the observations of the magnetic field map in c) are visible as individual nanotube fibres directed at various angles away from the dominant orientation of the fibres prescribing the main current flow.(c) Maps of the magnetic field arising from currents of 50 mA (top) and 100 mA (bottom) being passed through the carbon nanotube sample, measured by BEC-M. The trap shape (harmonic oscillation frequencies of $\qty{20}{\hertz}$ and $\qty{650}{\hertz}$ in the longitudinal and radial directions, respectively), and the atom-surface distance ($\qty{7.8}{\micro\metre}$) were kept constant throughout the measurement.
		}
		\label{fig:BECImplementation}
\end{figure*}
In this section, we describe our BEC-M system that provides the capability to perform microscopy on a series of samples, which is beneficial when aiming for a systematic comparison. We also demonstrate the application of the system to obtain experimental data from nanostructured samples.

The creation of the atomic probe for BEC-M requires both a UHV environment and optical access during various stages of the experiment (such as laser cooling and imaging). We have therefore implemented a scheme where initial trapping and probing take place in physically distinct regions, connected by atomic transport, on a single PCB (Figure~\ref{fig:BECImplementation}a). This provides versatility on the optical properties of the samples, design flexibility for the imaging system, and permits the use of a series of samples.

Certain types of samples can be placed directly in the central atom trapping region, making the system and its optimization procedure simpler. To demonstrate the feasibility of BEC microscopy over nanostructured samples, we have fabricated a percolating network of carbon nanotubes (CNT) on a substrate in the central region. Even though this sample is neither transparent nor optically flat, we empirically found it possible to create an efficient magneto-optical trap suitable for BEC production, illustrating the general robustness of the method.

We have scanned the probe with and without current running through the CNT sample and the difference is used to calculate the magnetic field generated by the sample as shown in Figure~\ref{fig:BECImplementation}c. For higher current (50~mA in the upper, 100~mA in the lower panel), the redistribution of atoms in the trap, and therefore the magnetic field detected, is more pronounced. The structure seen in the magnetic field component parallel to the atomic probe arises solely due to current components in the orthogonal direction in the sample. Although the CNT sample is aligned approximately parallel with the probe (along the $x$ direction), off-axis current flow (i.e., deviations into the $y$ direction) accounts for the BEC-M observations. Such structural features are typical in the CNT sample, as can be seen in Figure~\ref{fig:BECImplementation}b, which shows an AFM topography map of a representative region of the CNT sample.

\paragraph{Conclusion} 
We have experimentally demonstrated, and further analyzed through numerical studies, that the BEC-M technique is able to map current patterns at very low current levels---in the nanoampere range---where the sensitivity of thermal imaging methods is not sufficient. 

Our current density characterization will for the first time avoid assumptions regarding the link between thermal and conductive properties. Note here the importance of independently assessing current and temperature changes when their relationship is nonlinear. Combined with existing thermal imaging methods \cite{Estrada2011,Maize2015,Sannicolo2016} this relationship can be studied extensively, including modeling and testing individual junction behaviour and hot-spot formation in networks. 

At high currents where network failure is likely, the microsecond scale imaging duration of BEC-M allows for stroboscopic observation of the network dynamics leading to failure. 
Insights into the evolution of irreversible mechanisms in the networks will enable a better understanding of material and device failure and therefore aid the development of devices with enhanced macroscopic performance. 
Beyond current density mapping the BEC-M also allows for studies of atom-surface interactions \cite{Bender2014} which may be of great interest for sensing applications with nano-structures. 

\paragraph{Acknowledgement}
The authors gratefully acknowledge financial support through the Strategic Development Fund of the University of Sussex and the UK National Quantum Technology Hub in Sensing and Timing (EPSRC Reference EP/T001046/1). 
We thank Alan Dalton and Matthew Large for valuable discussions, Manoj Tripathi for providing AFM data, and Alice King and Christopher Brown for their help in fabricating samples.

\bibliography{BEC-M_bibtex}

\end{document}


\maketitle

\setstretch{1.1}

\section{BEC-M Measurement Procedure}

Here we describe in detail the BEC-M measurement procedure that was used for the experimental data presented in Figure 1 of the main text. First, an ultracold gas of rubidium-87 atoms in an ultra-high vaccuum ($10^{-10}$ mbar) environment was  prepared at a temperature of ~1uK using standard techniques of laser cooling, magnetic trapping, and forced evaporative cooling \cite{metcalf1999laser}. The atoms were then trapped by the atom chip \cite{Reichel_2011}, which is comprised of a variety of planar conductors microfabricated using lithography and gold evaporation. Precisely controlled currents in the atom chip conductors produce a magnetic field of desired spatial distribution.  Since atoms in correctly prepared (low-field seeking) magnetic quantum states are trapped in the regions of the lowest magnetic potential (proportional to the field magnitude), this allows for flexible manipulation of the atomic probe to vary its shape and position in three-dimensional space. Typically, currents of 300~mA to 1~A are passed through the atom chip conductors. After trapping, a further cooling phase brings the cloud below the transition temperature of 450~nK, producing a Bose-Einstein condensate (BEC) of approximately 100,000 atoms at a temperature of 200~nK. The resulting gas of atoms forms the atomic probe for subsequent microscopy measurements.

For the measurement presented in Figure 1 of the main text, the BEC is trapped by a conductor that is $200~\mu$m wide and $2~\mu$m thick, at an atom-surface distance of $1.4~\mu$m. To perform microscopy, the probe is then scanned across the surface of the conductor in the y-direction over a total distance of $80~\mu$m in $4~\mu$m steps, as shown in Figure~\ref{fig:SuppFig1} below. This translational movement of the probe is achieved using an additional vertical external bias magnetic field. At each position, the 2D atomic column density distribution is recorded using standard absorption imaging techniques \cite{Smith11}. A brief (1~ms) period of free expansion is implemented between extinguishing all magnetic fields and capturing each absorption image, which simplifies the quantitative analysis of the images. When scanning the atomic probe, meticulous attention was given to maintaining the atom-surface distance ($1.4~\mu$m), the trap shape (22~Hz and 1.4~kHz in the longitudinal and radial directions, respectively), and field minimum ($130~\mu$T) throughout.
\begin{figure*}[!ht]
		\centering
		\includegraphics[width=0.7\textwidth]{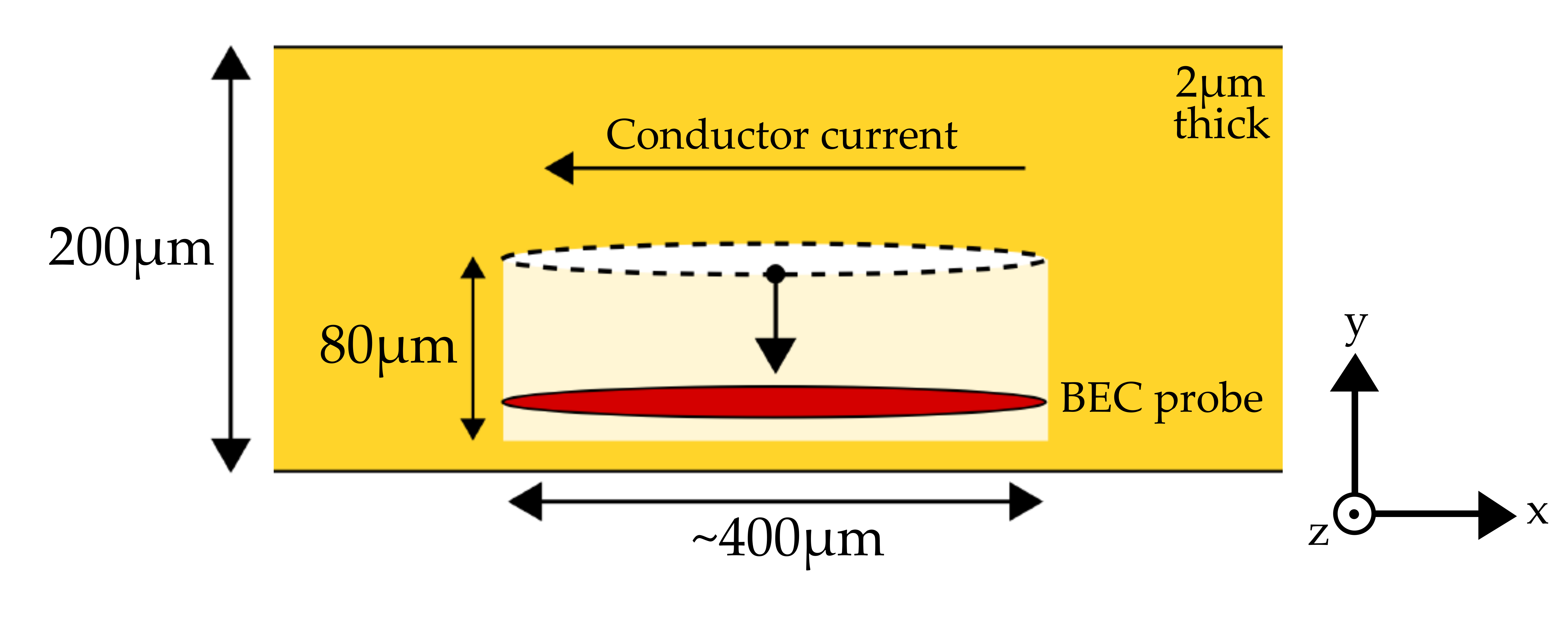}
		\caption{Schematic of the gold track used in the BEC-M demonstration. The atomic probe (red) was scanned over the surface in the light yellow region.}
		\label{fig:SuppFig1}
\end{figure*}

The measured 2D atomic column density distributions $n_{2\mathrm{D}}(x,z)$ are then converted into a series of 1D line densities $n_{1\mathrm{D}}(x)$, by integrating the signal along the $z$ direction. Each line density profile then corresponds to a different $y$ position, ultimately resulting in a 2D map in the $x-y$ plane. The deviations in atomic density in this map can be converted into a change in magnetic field component $B_x(x,y)$ using the approximation $ 2h \nu_{\perp}a_{sc}/\mu_{\mathrm{B}}\cdot\delta n_{1\mathrm{D}}(x)$, which is valid in the low density limit  ($n_{1\mathrm{D}}<\qty{100}{atoms \per \micro\metre}$) \cite{Gerbier2004}. Here $h$ is Planck's constant, $\nu_{\perp}$ is the radial trapping frequency, $a_{sc}$ is the s-wave scattering length, and $\mu_{\mathrm{B}}$ is the Bohr magneton. This magnetic field map is shown in Figure 1d of the main text.

The current density $J_y$ that gives rise to the $B_x$ component of the magnetic field can then be recovered from the measured field by using an inverse method based on the Green's function for the Biot-Savart law \cite{Roth1989}. This method gives a unique solution with the assumption that the source current is restricted to 2D (i.e.~does not have any variation along the $z$ direction), which is a resonable assumption in this case. This results in the current density map shown in Figure 1e of the main text.

\section{Nonlinear Thermo-Electric Model}

Here we provide further details of the model used in Figure 4 of the main text, which investigates the thermo-electric interplay in a nanowire network. The model demonstrates a simple example of how current path reconfiguration in a network can be understood via Joule heating of the wires and a nonlinear junction response to the temperature.

\begin{figure*}[!ht]
		\centering
		\includegraphics[width=0.8\textwidth]{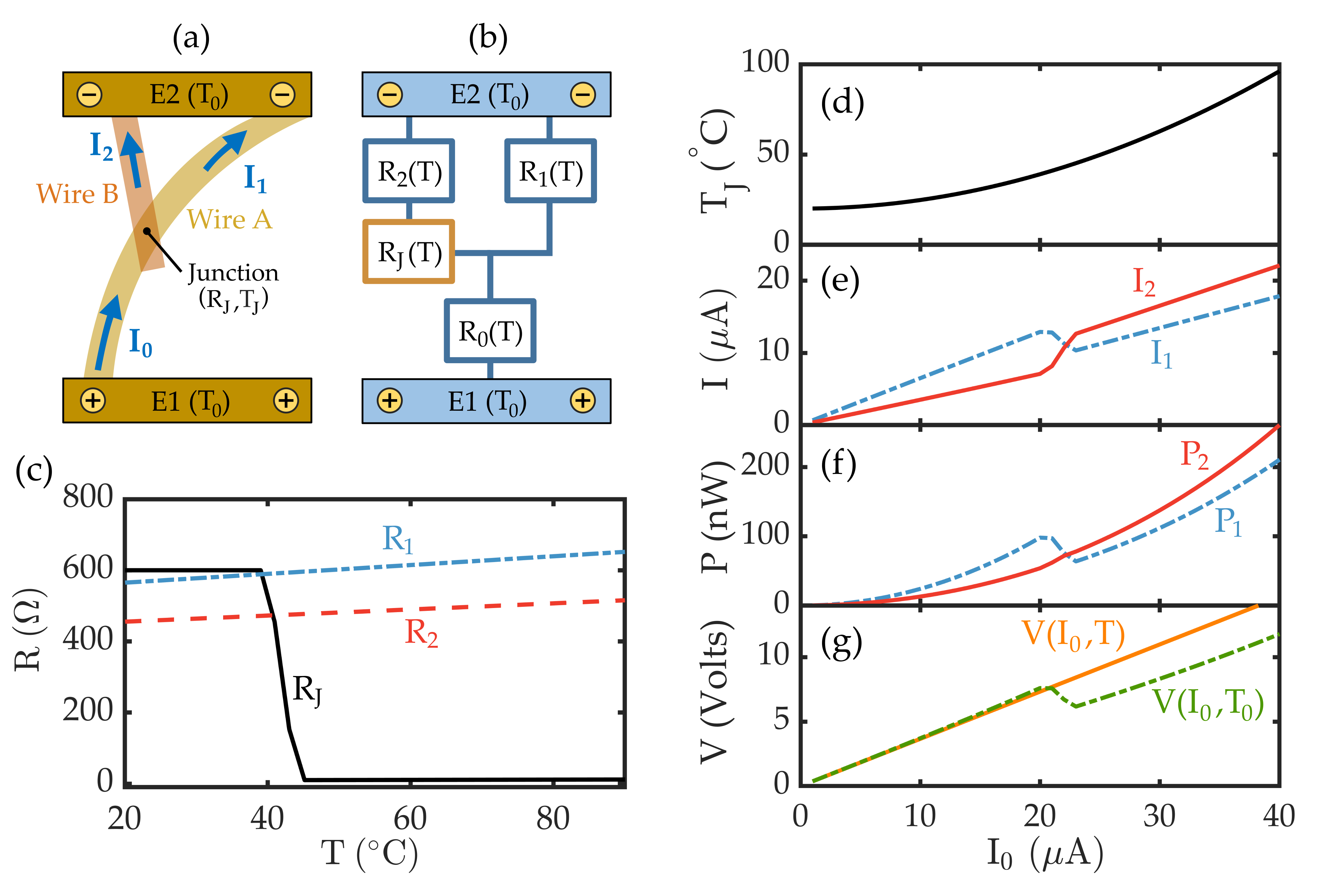}
		\caption{(a) Schematic of the model setup, showing two wires crossing at a junction between two electrodes. (b) Equivalent electrical circuit for the model, with each wire section having a temperature-dependent resistance. (c) Temperature-dependence of the wire sections and junction resistances. (d) Increasing junction temperature as a function of input current $I_0$. (e) Currents I$_{1}$ and I$_{2}$ and (f) power dissipation P$_{1}$ and P$_{2}$ in the two upper wire sections of the circuit. (g) Voltage across the circuit, showing the I-V characteristic of the circuit when taking into account the temperature dependence in the model $V(I_0,T)$, and when ignoring it $V(I_0,T_0)$.}
		\label{fig:SuppFig2}
\end{figure*}
The system considered is shown schematically in Figure 2a below, and consists of two wires (Wire A and Wire B) which overlap at a connecting junction, J. The variable input current $I_0$ is prescribed to flow from the positive electrode (E1) to the negative electrode (E2), and the relative currents $I_0,I_1$ and $I_2$ through each section will be distributed through the network according to the relevant resistances. The equivalent electrical resistance circuit for the model is shown schematically in Figure 2b. The lower and upper sections of Wire A have resistances $R_0$ and $R_1$, respectively, while Wire B has resistance $R_2$.

To include thermal effects in the model, we begin by prescribing a temperature-dependent resistance $R_{\mathrm{J}}(T_{\mathrm{J}})$ for the junction itself. We choose an abrupt drop in resistance at the junction from $\qty{600}{\ohm}$ to $\qty{10}{\ohm}$ as the temperature rises from $\qty{40}{\celsius}$ to $\qty{44}{\celsius}$. In the absence of extensive experimental characterization of $R_{\mathrm{J}}(T_{\mathrm{J}})$, we base our model on this assumption, as it is consistent with the measurements of Bellew et al.\cite{Bellew2015} This resistance characteristic is displayed in Figure 2c (solid line). Such a sudden drop in resistance could be a result of enhanced electrical contact due to material softening during heating, for example. 

In addition to the junction, we also specify the resistances $R_{i}(T)$ of the remaining wire sections according to their geometry. For the model discussed here, we choose the material of the wires to be silver, and use the bulk resistivity value of $\SI{1.59}{\micro \ohm \centi \meter}$. The silver nanowires are of circular cross-section \SI{60}{\nano\meter} in diameter, with lengths of \SI{200}{\micro\meter}, \SI{100}{\micro\meter}, and \SI{80}{\micro\meter} corresponding the sections $R_0,R_1$, and $R_2$, respectively. The resistivity temperature coefficient of bulk silver of \SI{0.0038}{\per\kelvin} was used for all wire sections, and the resulting temperature dependence of the resistances $R_1(T)$ and $R_2(T)$ is shown in Figure 2c (dashed lines).

To solve the model, we allow a fixed current $I_0$ between the electrodes E1 and E2. After numerically discretizing all wires into small segments, we calculate the temperatures and resistances in each segment along the wires from the steady-state solution of the 1D heat diffusion equation. As the heat source, we use Joule heating due to the variable power dissipation in the conductors, and for the thermal boundary conditions we fix the temperature of the extreme wire ends at $T_0$=\SI{20}{\celsius}, so that the electrodes act as thermal baths. In this case, the steady state situation admits an analytic solution, for which the resulting temperature distribution along a wire is parabolic $T(x) = \gamma/2\cdot x^2 -\gamma L/2\cdot x + T_0$. Here, $\gamma = I^2R/\kappa V$, where $\kappa$ is the thermal conductivity (419 W/m$\cdot$K for silver), and $V$ is the volume of the conductor. As the temperature rises due to power dissipation, the position-dependent resistances are then self-consistently adjusted accordingly. The junction then takes up the temperature of the local wire segment at all times.

The results of the model are shown in Figure~2d--g above. As the input current $I_0$ is increased, the temperature gradually rises due to power dissipation (Figure~2d). The ratio of currents in the two wire sections between E2 and the junction depends on $I_0$. For low values of $I_0$, Wire B carries less current than the upper section of Wire A ($I_2<I_1$, shown in Figure~2e), due to the large junction resistance. However, for larger currents ($I_0\sim\qty{20}{\micro\ampere}$) the temperature rises to the threshold value of $\sim$\SI{40}{\celsius}, at which point the drop in junction resistance is induced. The nonlinear junction response reconfigures the current path, and $I_2>I_1$.

This reconfiguration brings the system into the regime of super-Joule heating. Indeed, in contrast to the Joule heating power ($P_1=R_1I_1^2$) that scales with $I_1^2$, the power $P_2 = (R_2 + R_\mathrm{J})I_2^2$ exhibits a qualitatively different behaviour compared to $I_2^2$ (as shown in  Figure \ref{fig:SuppFig2}f). Experimentally, thermal imaging provides maps of the Joule heating power, but it does not give access to resistance or current information in the network directly. By obtaining the active current maps by BEC-M, the complete information necessary to model the percolating network and to work out individual junction behaviour would be available from the combination of the two methods. 
The reconfiguration of the current paths, however, can be quantitatively measured at various current levels over the entire network using the BEC-M alone, even at current levels as low as nanoamperes.

The voltage across the circuit [$V(I_0,T)$ in Figure \ref{fig:SuppFig2}g] shows a similar behaviour to that observed in Bellew et al. \cite{Bellew2015}. The linear $I$-$V$ (Ohmic) regime at low currents and a slight deviation from it at medium current levels is followed by a significant drop after which linearity is regained at larger currents. Such behaviour is considerably different from the case where temperature dependence is not taken into account [$V(I_0,T_0)$ in Figure \ref{fig:SuppFig2}g].

\section{CNT Measurement Details}

Here we give specific experimental details regarding the BEC-M measurement over a carbon nanotube (CNT) sample, which was presented in Figure~5 of the main text. 

The CNT sample was fabricated on a sample
strip, with dimensions \SI{18}{\milli\meter} long and \SI{3.5}{\milli\meter} wide. The sample strip was situated on a thin (\SI{110}{\micro\meter}) substrate, that additionally serves as quarter waveplate for controlling laser polarisation during the initial atom trapping.

For these data, the atomic BEC probe consisted of $\sim$40,000 atoms at a temperature of $\qty{400}{\nano\kelvin}$. The probe was oriented with the elongated axis along the $x$-direction, and then scanned across the current-carrying CNT sample surface over a range of $\qty{40}{\micro\metre}$ in steps of $\qty{780}{\nano\metre}$ along the $y$-direction. Throughout the scan, the trap shape (a harmonic trap with frequencies of \SI{20}{\hertz} and \SI{650}{\hertz} in the longitudinal and radial
directions, respectively), and the atom-surface distance of \SI{7.8}{\micro\meter} were kept constant.

The atomic distributions were recorded at each position, and the 2D map of 1D atomic line densities as a function of $y$-position was constructed (by integrating the column density images as described in the \textit{BEC-M Measurement Procedure} section described above). We have recorded atomic probe scans under three current conditions: 1) without any current running through the CNT sample as a reference, 2) an identical scan with $\qty{50}{\milli\ampere}$ flowing through the sample, 3) again with $\qty{100}{\milli\ampere}$ through the sample. All data are compiled from three repeats of each to demonstrate reproducibility and increase the signal-to-noise ratio. The atomic density scans are shown in Figure~3 below. With higher current, the redistribution of atoms in the trap is even more pronounced.

The difference between the non-zero current and the reference scan for each of the two 50mA and 100mA cases were then used to calculate the additional magnetic field component generated by the sample. These final magnetic field scans are shown in Figure~5 of the main text.  
\begin{figure*}[!ht]
		\centering
		\includegraphics[width=0.6\textwidth]{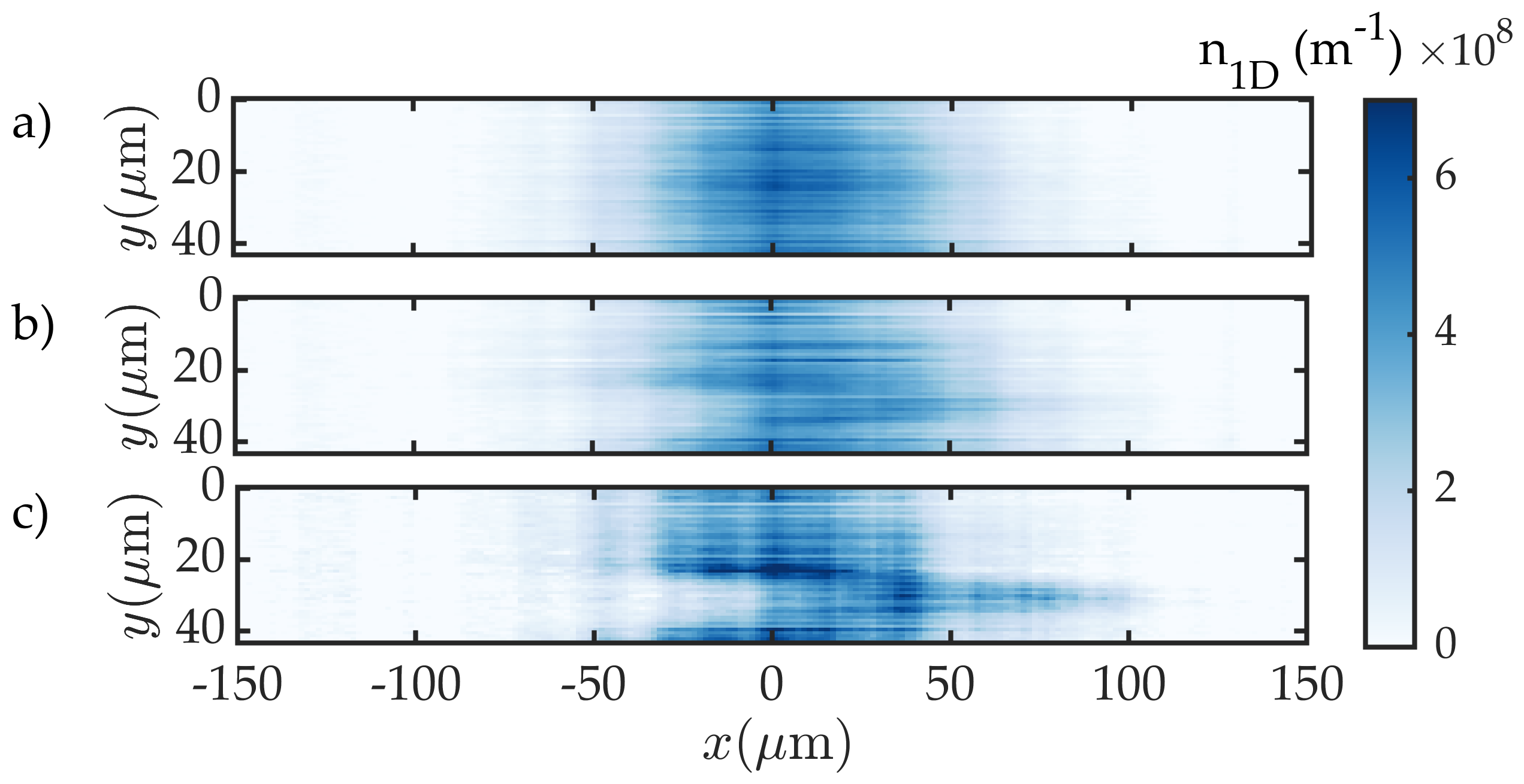}
		\caption{Measured atomic line density scans for varying currents through the carbon nanotube sample. (a)~No current, for reference. (b) and (c) \SI{50}{\milli\ampere}  and \SI{100}{\milli\ampere} through the sample, respectively.}
		\label{fig:SuppFig3}
\end{figure*}

\printbibliography[
heading=bibintoc,
title={References}
]